\documentstyle[amssymb,twocolumn,aps,prb,epsf]{revtex}

\begin{document}
\draft
\title{Mesoscopic Kondo screening effect in a single-electron transistor embedded
in a metallic ring}
\author{Hui Hu$^1$, Guang-Ming Zhang$^{1,2}$, Lu Yu$^{3,4}$}
\address{$^1$Department of Physics, Tsinghua University, Beijing 100084, China\\
$^2$Center for Advanced Study, Tsinghua University, Beijing 100084, China\\
$^3$Abdus Salam International Center for Theoretical Physics, P. O. Box 586,%
\\
Trieste 34100, Italy\\
$^4$Institute of Theoretical Physics, Academic Sinica, Beijing 100080, China}
\date{\today}

\twocolumn[\hsize\textwidth\columnwidth\hsize\csname@twocolumnfalse\endcsname

\maketitle

\begin{abstract}
We study the Kondo screening effect generated by a single-electron transistor
or quantum dot embedded in a small metallic ring. When the ring circumference
$L$ becomes comparable  to the fundamental length scale
$\xi _K^0=\hbar \upsilon _F/T_K^0$ associated with the {\it bulk} Kondo temperature,
 the Kondo resonance is strongly affected,  depending on  the total
 number of electrons ({\it modulo} 4) and magnetic flux threading the ring.
 The resulting Kondo-assisted persistent currents are also calculated
 in both Kondo and mixed valence regimes, and the maximum values are
 found in  the crossover region.
 \end{abstract}


 ]

The effect of magnetic impurities on metals--- the Kondo problem has been
studied for nearly half a century \cite{hewson}, and attracts continued
interest till now in attempts to understand heavy fermion materials and
high-T$_{c}$ superconductors. Recent progress in the nano-fabrication
technique of electronic devices has enabled investigation of the Kondo
effect by means of single-electron transistors (SETs) or quantum dots (QDs)
in a controllable way \cite{qdots,qdot,qd,dgg,ring-dot,jiyang}. When it
contains an odd number of localized electrons, a QD may be described by the
Anderson impurity model, exhibiting the Kondo effect characterized by the
appearance of a sharp resonance near the Fermi energy\cite{glazman,nglee},
and it provides a new channel for electric current flowing through the QD.

To investigate the quantum phase interference, a device setup of a QD
embedded in a mesoscopic metallic ring has been proposed\cite
{zvyagin,buttiker,ferrari,kang,affleck}. In such a geometry the screening
cloud is trapped in the ring, and one can measure the persistent current
(PC) in the ring induced by the external magnetic flux to understand the
coherent transport through the dot without attaching leads to it. Based on
perturbation calculations, Affleck and Simon \cite{affleck} recently pointed
out that in the Kondo regime, the PC is a universal scaling function of $\xi
_K^0/L$, where $\xi _K^0=\hbar \upsilon _F/T_K^0$ is the characteristic
length scale associated with the {\it bulk} Kondo temperature $T_K^0$, and
crosses over from the value for a perfect ring with no QD in the limit $\xi
_K^0/L\ll 1$ to a vanishing small magnitude for $\xi _K^0/L\gg 1$, in
contrast to the variational calculations \cite{kang}.

In this Letter, we would like to further clarify the underlying physics
within a model of an Anderson magnetic impurity embedded in a mesoscopic
metallic ring. Explicitly considering the discrete spectrum of conduction
electrons in the ring using the slave boson mean field (SBMF) approach, we
show that the Kondo resonance is strongly dependent on the ratio of ring
circumference $L$ to the screening cloud size $\xi _{K}^{0}$ ($L=N$, as the
lattice spacing is assumed to be unity), the total number of electrons ({\em %
\ modulo} $4$), and the magnetic flux threading the ring. The resulting
Kondo-assisted PCs are calculated in both Kondo and mixed-valence regimes.
Our numerical results are in good agreement with those of perturbation
calculations in the limits of $\xi _{K}^{0}/L\ll 1$ and $\xi _{K}^{0}/L\gg 1$
\cite{affleck}. The {\em mesoscopic} Kondo effect was also studied in
ultrasmall metallic grains \cite{delft}.

{\it SBMF approach.} --- A QD embedded in a mesoscopic Aharonov-Bohm (AB)
ring can be described by a one-dimensional tight-binding Hamiltonian with $N$
lattice sites including an Anderson impurity at site ``0''. The model
Hamiltonian is given by:

\begin{eqnarray}
&&H=-t\sum\limits_{j=1}^{N-2}\sum\limits_\sigma \left( c_{j\sigma }^{\dagger
}c_{j+1\sigma }+h.c.\right) +Ud_{\uparrow }^{\dagger }d_{\uparrow
}d_{\downarrow }^{\dagger }d_{\downarrow }  \nonumber \\
&&+\sum\limits_\sigma \left[ \varepsilon _dd_\sigma ^{\dagger }d_\sigma
-\left( t_Ld_\sigma ^{\dagger }c_{1\sigma }+t_Re^{i\phi }c_{N-1\sigma
}^{\dagger }d_\sigma +h.c.\right) \right] .
\end{eqnarray}
Here $t_L$ ($t_R$) denotes the hopping between the QD and the left (right)
neighboring site of the ring. The phase factor $\phi $ is defined by $2\pi
\Phi /\Phi _0$, where $\Phi $ and $\Phi _0=h/e$ are external magnetic flux
and the flux quantum, respectively. We only consider a single
spin-degenerate energy level $\varepsilon _d$ of the QD. Since the Coulomb
repulsion $U$ is generally large in experiments, the limit of $U\rightarrow
\infty $ can be taken. A parameter describing the coupling strength between
the QD and the ring is introduced $\Gamma =\pi \rho (\varepsilon _F)\left|
t(\varepsilon _F)\right| ^2$, where $\rho (\varepsilon _F)$ and $%
t(\varepsilon _F)$ represent the ring density of states (DOS) and hopping at
the Fermi energy $\varepsilon _F$, respectively. In this paper, only the
half-filled case ($N_e=N$, where $N_e$ is the total number of localized and
delocalized electrons) is considered. In the thermodynamic limit, we would
have $\varepsilon _F=0$, $\rho (0)=N/(2\pi t)$, $t^2(0)=2(t_L^2+t_R^2)/N$,
and $\Gamma =(t_L^2+t_R^2)/t$.

The strong coupling SBMF theory has been well-known to be a good
approximation in describing the Kondo physics at low temperatures. Here we
will apply this method to the Anderson impurity model in the presence of a
magnetic flux, and we believe that it can still yield reliable results.
Because of the infinite $U$ limit, the impurity operator can be expressed as 
$d_\sigma ^{\dagger }=f_\sigma ^{\dagger }b$ in the slave-boson
representation, where the fermion $f_\sigma $ and the boson $b$ describe the
singly occupied electron and hole states, respectively. The constraint $b$%
{\bf $^{\dagger }$}$b+\sum_\sigma f_\sigma ^{\dagger }f_\sigma =1$ has to be
imposed. When the mean field (MF) approximation is made, the boson operators
are replaced by a c-number $b_0$, and the constraint is implemented by
introducing a chemical potential $\lambda _0$. Therefore, the MF Hamiltonian
is written as: 
\begin{eqnarray}
&&H_{mf}=-t\sum\limits_{j=1}^{N-2}\sum\limits_\sigma \left( c_{j\sigma
}^{\dagger }c_{j+1\sigma }+h.c.\right) +\sum\limits_\sigma \tilde{\varepsilon%
}_df_\sigma ^{\dagger }f_\sigma  \nonumber \\
&&-\sum\limits_\sigma \left( \tilde{t}_Lf_\sigma ^{\dagger }c_{1\sigma }+%
\tilde{t}_Re^{i\phi }c_{N-1\sigma }^{\dagger }f_\sigma +h.c.\right) +\lambda
_0(b_0^2-1),  \label{hmf}
\end{eqnarray}
which is essentially a noninteracting system with a set of renormalized
parameters $\tilde{\varepsilon}_d=\varepsilon _d+\lambda _0$ and $\tilde{t}%
_{L(R)}=b_0t_{L(R)}$. Since the conduction electrons themselves do not form
a {\it complete} set, the Fourier transformation can not be used to
diagonalize $H_{mf}$. As a result, the ground state energy $E_{gs}$ can only
be evaluated by exact diagonalization as 
\begin{equation}
E_{gs}=\sum\limits_{m\sigma }^{occ}\varepsilon _{m\sigma }(\lambda
_0,b_0)+\lambda _0(b_0^2-1),
\end{equation}
where the summation over $m$ includes all occupied levels of $H_{mf}$. The
values of $\lambda _0$ and $b_0$ can be determined self-consistently by
minimizing $E_{gs}$ with respect to $\lambda _0$ and $b_0$. Once the
solutions $\lambda _0^{*}$, $b_0^{*}$ and $E_{gs}=E_{gs}^{*}(\lambda
_0^{*},b_0^{*})$ are obtained, one can calculate the conduction electron DOS
at site $N_i$ by

\[
\rho _{N_{i}\sigma }(\omega )=-\frac{1}{\pi }{\rm 
\mathop{\rm Im}%
}\text{ }\left\langle 0\left| c_{N_{i}\sigma }(\omega +i\gamma
-H_{mf})^{-1}c_{N_{i}\sigma }^{\dagger }\right| 0\right\rangle , 
\]
and the PC: $I=-\frac{e}{\hbar }\frac{\partial E_{gs}}{\partial \phi }$,
where $\left| 0\right\rangle $ represents the vacuum state, and $\gamma $ is
introduced to denote the half-maximum width of the spectrum lines in the
DOS. Note that the DOS of QD is given by $\rho _{QD}(\omega )=b_{0}^{2}\rho
_{0\sigma }(\omega )$, and the ``Kondo correlation energy'' $T_{K}$ for the
finite-size systems can be defined by the energy gain due to coupling
between the QD and the ring: 
\begin{equation}
T_{K}=\varepsilon _{d}-\varepsilon _{F}-(E_{gs}-E_{gs}^{0}),
\end{equation}
where $E_{gs}^{0}$ denotes the ground state energy of the noninteracting
Hamiltonian $H_{0}$ with electron number $N_{e}$ and an {\it open} boundary
condition, and $\varepsilon _{F}$ is the energy of the highest occupied
level. $T_{K}$ is thus the Kondo temperature in the finite size system, and
it will approach $T_{K}^{0}$ in the thermodynamical limit ($L\rightarrow
\infty $). In this sense, we will refer to $T_{K}$ as a {\it mesoscopic}
Kondo temperature.

The Anderson impurity model in QD realizations also exhibits distinct
behaviors in different regimes \cite{dgg}. Here we will mainly focus on the
Kondo regime, namely, the highest occupied energy level of the QD has a
single occupancy. As a typical case, we choose $\varepsilon _{d}=-0.7t$, $%
t_{L}=t_{R}=0.35t$, and $\Gamma =0.245t$, corresponding to $%
T_{K}^{0}=0.0105t $ or $\xi _{K}^{0}=191$ in the thermodynamic limit.

{\it DOS of QD. }--- In the absence of the external magnetic flux, the DOS
of QD $\rho _{QD}(\omega )$ is obtained by numerically solving the SBMF
self-consistent equations for various $\xi _{K}^{0}/L$ and are summarized in
Fig. 1, where the half maximum width of the spectrum lines has been chosen
as $\gamma =0.5T_{K}^{0}$, and different choices of $\gamma $ will not
change the results much. For $\xi _{K}^{0}\ll L$, the shape of the Kondo
resonance is indistinguishable from the bulk case. When $L$ is decreased
down to $\xi _{K}^{0}$, however, it splits up into a set of subpeaks, in a
way which depends strongly on the total numbers of lattice sites $N$ ({\em %
modulo} $4$). The case of $N=4n+2$ shows two main peaks with nearly the same
weight around $\varepsilon _{F}$, whereas the others only have a single main
peak at slightly different positions. Despite developing subpeaks, the Kondo
resonance retains its basic feature, {\em i.e.}, the profile of the DOS
around $\varepsilon _{F}$ is preserved within a few $T_{K}^{0}$. In other
words, the Kondo correlations induced by spin-flip processes between the
singly occupied electron on the QD and the conduction electrons persist even
in a very small system.

Features for the $\xi _{K}^{0}\sim L$ case are {\it mainly} determined by
the highest occupied and lowest unoccupied single particle energy levels of $%
{\cal H}_{mf}$. For $N=4n$, when the localized electron on QD binds a
delocalized electron at the energy range $\left| \varepsilon \right|
\lesssim \Gamma $ to form a Kondo singlet near $\epsilon _{F}=0$, the energy
difference between the highest occupied and lowest unoccupied levels is very
small, and a single peak is dominant in the DOS. In contrast, for $N=4n+2$,
there is a large difference in the resulting highest occupied and lowest
unoccupied energy levels, so the Kondo resonance peaks appear both below and
above $\varepsilon _{F}$ with almost the same weight. On the other hand, for
the odd parity cases of $N=4n\pm 1$, the highest energy levels are singly
occupied below and above $\varepsilon _{F}$, respectively, and a single
Kondo resonance thus appears below and above $\varepsilon _{F}$ separately.
This feature may have significant consequences.

The DOS of the QD is also strongly affected by the external magnetic flux.
This is illustrated in Fig. 2, which shows the $\phi $ dependence of $\rho
_{QD}(\omega )$ for $\xi _{K}^{0}/L=1.0$. A striking feature is that the DOS 
{\it with }$N${\it \ sites and zero flux is exactly\ equal to the one with }$%
N+2${\it \ sites and a half flux quantum}. Such a feature is related to the
nature of the strong coupling fixed point for the model, where one electron
is trapped in the symmetric sites around the impurity and other delocalized
conduction electrons gain a resonant phase shift from the scattering off the
Kondo singlet. It is therefore sufficient to consider only the case of $N=4n$
and $4n+1$. For even parity of $N=4n$, the single main peak in the DOS
gradually splits into two symmetric peaks as $\phi $ increases to $\pi $,
and they recombine again into a single peak as $\phi $ is further increased
to $2\pi .$ For odd parity of $N=4n+1$, the situation is quite different. A
single main peak in DOS persists asymmetrically in the whole range of $\phi $%
, and its position is shifted with a period of $2\pi $. However, in this odd
parity cases there is an important symmetry in the DOS of QD: $\rho
_{QD}(-\omega ,\phi )=\rho _{QD}(\omega ,\pi +\phi )$, leading to the period
halving in the PCs.

{\it Kondo screening cloud. }--- In order to determine the size of the Kondo
screening cloud, Fig. 3a illustrates the DOS of the QD and selected lattice
sites for a very large ring $\xi _{K}^{0}/L=0.1$. At $\omega \approx 0$, the
DOS of lattice sites $N_{i}$ {\it near} the QD is strongly affected by the
Kondo resonance, and clearly exhibits a parity-dependence, {\it i.e}., it is
enhanced for $N_{i}=$ even and suppressed for $N_{i}=$ odd. However, such an
influence does weaken significantly once the lattice site is away from the
QD. This is illustrated in the change of DOS at $\varepsilon _{F}$, as shown
in Fig. 3b, which is obtained by substracting the background for the
noninteracting part $H_{0}$. We observe that $\Delta \rho _{N_{i}\sigma
}(\omega =0)$ drops {\it exponentially} with increasing $N_{i}$ as expected.
The decay length $\xi _{K}$ deduced from numerical fitting is roughly $190$,
which is close to the Kondo correlation length scale in the thermodynamic
limit $\xi _{K}^{0}=191$. Note that this is not an accurate procedure to
determine this characteristic length scale.

{\it PC}. --- The most convenient way to detect the above Kondo screening
cloud in experiments is to measure the PC in a ring induced by the magnetic
flux. Such measurements have been performed recently on micron size rings
not containing a QD \cite{chand} and slightly different experimental setup
with a QD \cite{jiyang}. In Fig. 4 we show the PC vs magnetic flux for $\xi
_{K}^{0}/L=0.3$ (dashed line) and $1.5$ (solid line). A striking feature can
be clearly seen $I^{N+2}(\phi )=I^{N}(\phi +\pi )$ due to the fact that the
structures of DOS of QD have an exact $\pi $ shift from the system with $N$
sites to that with $N+2$ sites. It is obvious that the behavior of PC
differs greatly in the limit $\xi _{K}^{0}/L\ll 1$ from that for $\xi
_{K}^{0}/L\sim 1$. In the former case the system resembles an idea metallic
ring, implying a high electron transmission through the QD. When $L$ becomes
comparable to $\xi _{K}^{0}$, the PC does weaken systematically and their
profiles are close to sinusoidal. Moreover, the PC for odd parity appears to
be much smaller than that for even parity, which also indicates a weakening
of the Kondo screening effect and the QD simply acts as an {\it incoherent}
scatter. Note that these features are in good agreement with the recent
perturbation theory results \cite{affleck}, but are partly in contradiction
with those of variational calculations \cite{kang}.

In order to fully display the dependence on the ring circumference, the PCs
and Kondo correlation energies $T_{K}$ are calculated as functions of $\xi
_{K}^{0}/L$ in the cases of $\phi =\pi /4$ and $\pi /2,$ displayed in Fig.5.
We find the PCs interpolate smoothly between the limits of small and large $%
\xi _{K}^{0}/L$. Both $I/I_{0}$ and $T_{K}/T_{K}^{0}$ curves follows {\it \
universal} scaling functions of the ratio $\xi _{K}^{0}/L$, depending on the
total number of electrons ({\it modulo} 4). It is rather surprising that the 
{\it mesoscopic Kondo temperatures }$T_{K}${\it \ in small rings are much
higher than the bulk value }$T_{K}^{0}$, implying that the corresponding
Kondo correlation lengths $\xi _{K}$ are much {\it \ smaller} than the bulk
value $\xi _{K}^{0}$.

Finally, we study the properties of the PCs for different occupations of the
highest occupied energy level of QD by tuning the dot energy level $%
\varepsilon _{d}$. In Fig. 6 the PC is shown as a function of $\varepsilon
_{d}$ for various $\xi _{K}^{0}/L$ at $\phi =\pi /4$. For different
parities, the PCs exhibit an asymmetric structure and their maxima appear
near the crossover boundary of the Kondo and mixed-valence regimes. In the
Kondo regime, the PCs are {\it suppressed} as $L$ decreases, indicating the
effects of weakening in the Kondo-assisted tunnelings, while in the
mixed-valence regime, however, the PCs are almost unchanged as $L$ varies.
These distinct features suggest that the tunneling mechanisms in the Kondo
and mixed valence regimes are quite different.

In conclusion, we have calculated the mesoscopic Kondo screening effect in a
single-electron transistor embedded in a metallic ring. The Kondo resonance
is found to be strongly dependent on the dimensionless ratio $\xi _{K}^{0}/L$%
, the total number of electrons ({\em modulo} $4$), and the external flux
threading the ring. The screening effects on the PCs are expected to be
observable in experiments.

H. Hu acknowledges the support of Profs. Jia-Lin Zhu and Jia-Jiong Xiong,
and a research grant from NSF-China (No. 19974019). G.-M. Zhang was
supported by NSF-China (Grant No. 10074036) and the Special Fund for Major
State Basic Research Projects of China (G2000067107).

\begin{center}
{\bf Figures Captions}
\end{center}

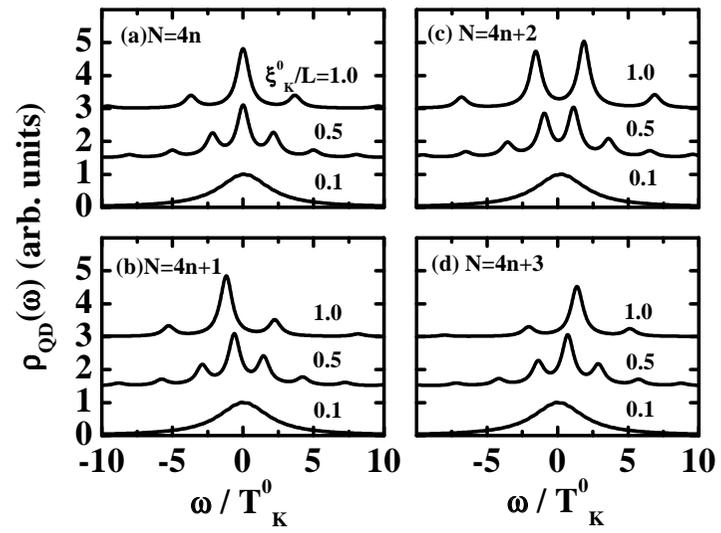
Fig.1. $\rho _{QD}\left( \omega \right) $ for various values of $\xi
_{K}^{0}/L$ in the absence of a magnetic flux. The number of lattice sites $%
N = L $ corresponds to (a) $4n$, (b) $4n+1$, (c) $4n+2$ and (d) $4n+3$,
respectively. Curves are offset by $1.5$ units each, and the half-maximum
width of the spectrum lines $\gamma =0.5T_{K}^{0}$. \newline

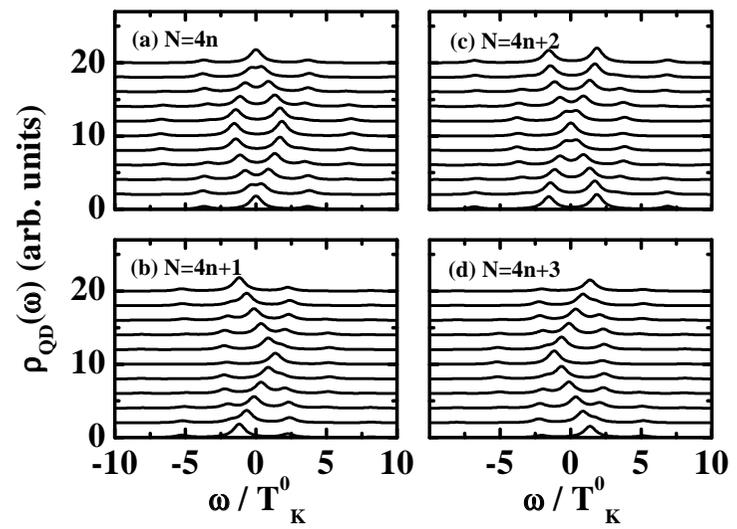
Fig.2. $\rho _{QD}\left( \omega \right) $ of $\xi _K^0/L=1.0$ for different
external flux. In each subplot, from the bottom up, $\phi =0\sim 2\pi $ is
increased at steps of $0.2\pi $. Curves are offset by $2$ units each.\newline

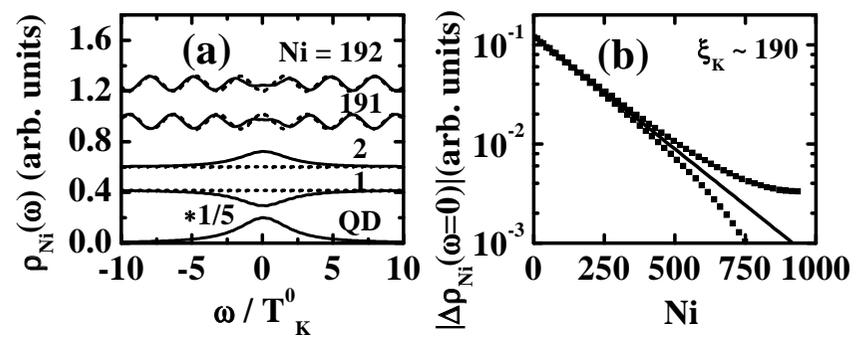
Fig.3. (a) $\rho _{QD}\left( \omega \right) $ and selected lattice sites $%
\rho _{N_{i}}(\omega )$ for $\xi _{K}^{0}/L=0.1$. The dashed lines denote
the corresponding DOS for $H_{0}$, and curves labeled by $N_{i}$ = $1,2,191,$
and $192$ are offset by $0.1,0.6,0.8,$ and $1.1$ units, respectively. The
curve for $\rho _{QD}\left( \omega \right) $ is reduced by five times. (b)
The change of DOS $\Delta \rho _{N_{i}\sigma }(\omega =0)$ is calculated by
substracting the background for $H_{0}$ as a function of the lattice site $%
N_{i}$ . The solid line fits the exponential decay and a characteristic
length is obtained $\xi _{K}\sim 190.\newline
$

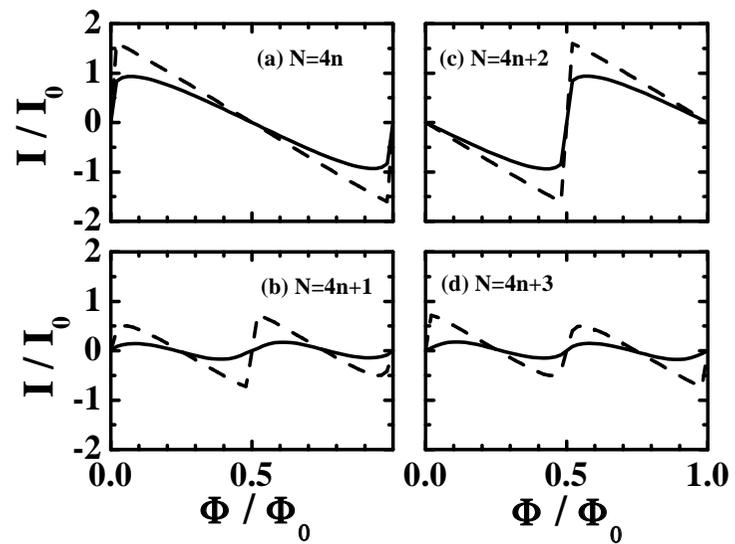
Fig.4. Persistent currents versus flux with the total number of lattice
sites $4n+i$ ($i=0,1,2,3$) for $\xi _{K}^{0}/L=0.3$ (dashed lines) and for $%
\xi _{K}^{0}/L=1.5$ (solid lines). $I_{0}$ denotes the corresponding PC of
an ideal metallic ring.\newline

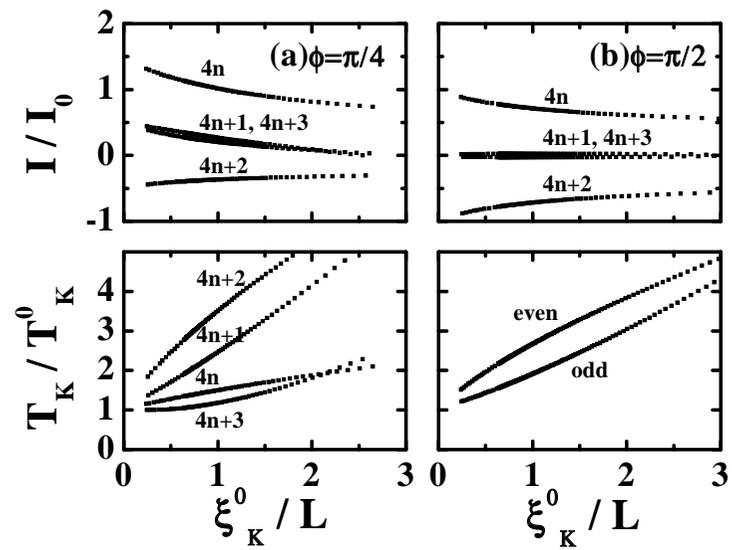
Fig.5. Persistent currents and Kondo temperature $T_{K}$ as functions of $%
\xi _{K}^{0}/L$ for $\phi =\pi /4$ and $\pi /2$. Both $I/I_{0}$ and $%
T_{K}/T_{K}^{0}$ curves follow universal scaling functions of the ratio $\xi
_{K}^{0}/L$.\newline

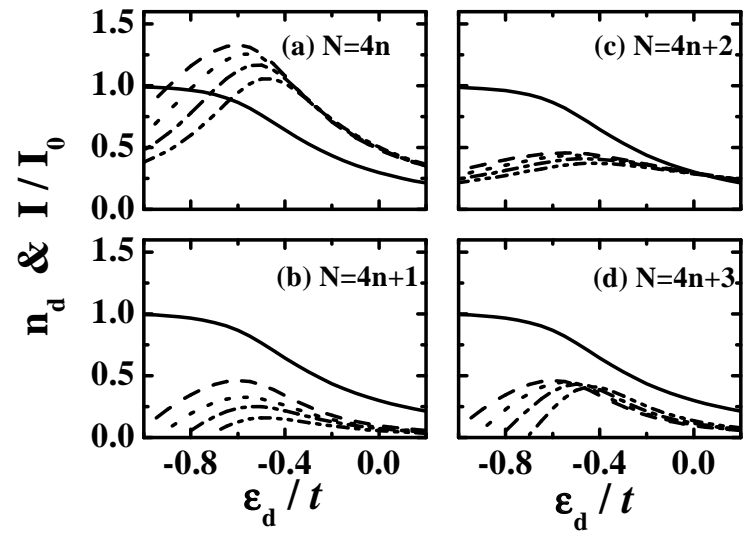
Fig.6. Occupation number of the highest occupied energy level of QD (solid
lines) $n_{d}$ and PCs $I/I_{0}$. The dashed, dotted, dash-dotted, and
dash-dot-dotted lines show the data for $\xi _{K}^{0}/L=0.3$, $0.6$, $1.2$,
and $2.4$, respectively, as functions of $\varepsilon _{d}$ for different
parities at $\phi =\pi /4$.

\end{document}